\documentclass[aps, twocolumn, letterpaper, superscriptaddress]{revtex4}

\usepackage{amsmath}
\usepackage{amssymb}
\usepackage{xspace}
\usepackage{bbold}
\usepackage{graphicx}

\usepackage{ifpdf}
\ifpdf
\fi

\newcommand{\eq}[1]{(\ref{#1})}
\newcommand{\Eq}[1]{Eq.~(\ref{#1})}
\newcommand{\Eqs}[1]{Eqs.~(\ref{#1})}
\newcommand{\Fig}[1]{Fig.~\ref{#1}}
\newcommand{\Sec}[1]{Sec.~\ref{#1}}
\newcommand{\Ref}[1]{Ref.~\cite{#1}}
\newcommand{\Refs}[1]{Refs.~\cite{#1}}

\newcommand{\eg}{{e.g.,\/}\xspace}
\newcommand{\ie}{{i.e.,\/}\xspace}

\newcommand{\trH}{H}
\newcommand{\ccdot}{}

\newcommand{\pd}{\partial}
\newcommand{\del}{\nabla}
\newcommand{\mc}[1]{\mathcal{#1}}

\renewcommand{\vec}[1]{{\boldsymbol{\rm #1}}}
\newcommand{\msection}[1]{{\it #1. -- }}

\begin{document}

\title{Mode conversion in cold low-density plasma with a sheared magnetic field}

\author{I.~Y. Dodin}
\affiliation{Princeton Plasma Physics Laboratory, Princeton, New Jersey 08543, USA} 
\author{D.~E. Ruiz}
\affiliation{Sandia National Laboratories, P.O. Box 5800, Albuquerque, New Mexico 87185, USA}
\author{S. Kubo}
\affiliation{National Institute for Fusion Science, National Institutes of National Sciences, 509-5292, Toki, Gifu, Japan}

\begin{abstract}
A theory is proposed that describes mutual conversion of two electromagnetic modes in cold low-density plasma, specifically, in the high-frequency limit where the ion response is negligible. In contrast to the classic (Landau--Zener-type) theory of mode conversion, the region of resonant coupling in low-density plasma is not necessarily narrow, so the coupling matrix cannot be approximated with its first-order Taylor expansion; also, the initial conditions are set up differently. For the case of strong magnetic shear, a simple method is identified for preparing a two-mode wave such that it transforms into a single-mode wave upon entering high-density plasma. The theory can be used for reduced modeling of wave-power input in fusion plasmas. In particular, applications are envisioned in stellarator research, where the mutual conversion of two electromagnetic modes near the plasma edge is a known issue.
\end{abstract}

\maketitle

\section{Introduction}

Mode conversion (MC) is the exchange of action (quanta) between normal modes of a dispersive medium when the parameters of the medium evolve in time or in space \cite{book:kravtsov, ref:zheleznyakov83, ref:bliokh01, book:tracy}. Here, we discuss linear MC, which is most efficient when both the frequencies and the wave vectors of interacting modes are close to each other. Regions of such resonant interaction are usually assumed well-localized, so MC theories typically approximate the coupling matrix with its Taylor expansion near the resonance. Then, the general two-wave coupling problem can be reduced \cite{book:tracy, ref:tracy03, ref:tracy93, ref:friedland87b, ref:friedland87}, at least in the absence of dissipation, to the classic Landau--Zener problem from quantum mechanics \cite{ref:landau32, ref:zener32}. This leads to compact asymptotic formulas for the mode amplitudes (which were also rediscovered \textit{ad~hoc} in various contexts; \eg see \Refs{ref:zheleznyakov83, book:ginzburg, tex:erokhin79, ref:kravtsov96}). However, there are systems where this somewhat universal ``Landau--Zener paradigm'' is inapplicable. They include inhomogeneous media with degenerate and nearly-degenerate wave spectra, such as isotropic or weakly anisotropic dielectrics \cite{ref:bliokh08, my:qdiel, ref:kravtsov96} and nonmagnetized or weakly-magnetized plasmas as a special case \cite{my:covar}. Although the procedure for deriving the governing equations for such media is known in general \cite{my:covar, phd:ruiz17}, calculating the coupling matrices and solving the wave equations explicitly remains an open research area.

Here, we study MC in a specific medium with a nearly-degenerate wave spectrum, namely, cold magnetized low-density plasma. As opposed to the standard treatment of the MC in magnetized plasma \cite{ref:preinhaelter73}, which is known as the \mbox{O-X} conversion, our theory allows for a sheared magnetic field. Previous theoretical studies of the \mbox{O-X} conversion in a sheared field \cite{ref:fidone71, ref:melrose74, ref:zheleznyakov79, ref:kocharovskii80, ref:brambilla87, ref:airoldi89, ref:bellotti99, tex:segre01, ref:popov10, ref:kubo15} were either pursued numerically or assumed planar geometry or specific limits (\eg the high-density limit), or addressed the dynamics of polarization instead of mode amplitudes \textit{per~se}. Hence, there is still a lack of a general analytical theory that could explicitly describe the exchange of quanta between the electromagnetic (EM) modes in the low-density case. 

The problem of MC in cold three-dimensional low-density plasma with a sheared magnetic field is presently of applied interest in stellarator research \cite{ref:kubo15, ref:notake05, ref:tsujimura15}. Due to a strong magnetic shear and a relatively smooth density profile near the plasma edge in a stellarator, an externally-launched single-mode EM wave can lose its quanta to the other EM mode through MC, hence affecting the overall wave-plasma coupling in the device. The MC occurs due to the fact that, in low-density plasma, both EM modes have close-to-vacuum dispersion, \ie are nearly-resonant; then, even a weak inhomogeneity of the magnetic field can couple them easily. To improve the efficiency of the wave-power input into a stellarator, a simple three-dimensional theory of MC in edge plasma with a sheared magnetic field could be beneficial. Here, we construct such theory analytically by considering the plasma density as a small parameter. 

The paper is organized as follows. In \Sec{sec:general}, we introduce a reduced equation for MC in a general context, basically, by restating results from \Refs{my:covar, my:qdirac, ref:friedland87}. Later, this formulation is tailored to weakly anisotropic media; namely, an ordinary differential equation (ODE) is derived for the mode complex envelope along geometrical-optics (GO) rays. The result represents an alternative to the well-known Budden-Kravtsov equations \cite{ref:zheleznyakov83, ref:kravtsov96}. In appropriate variables, the coefficients in our ODE depend on the medium parameters through only one real function. Moreover, if this functions remains smooth enough, only its asymptotics matter. Using this, we propose a simple method for predicting how the amplitudes of the two EM modes evolve within a given wave. In \Sec{sec:plasma}, we apply these findings to cold magnetized low-density plasma and explain how they can be used for optimizing the wave-power input in fusion plasmas. In \Sec{sec:conc}, we summarize our results. 

\section{General theory}
\label{sec:general}

\subsection{Basic equations}
\label{sec:basic}

Let us consider a stationary EM wave governed by
\begin{gather}\label{eq:weq}
\hat{\vec{D}} \ccdot \vec{E} = 0.
\end{gather}
The dispersion operator $\smash{\hat{\vec{D}}}$ that determines the evolution of the electric field $\vec{E}$
is obtained by combining Ampere's and Faraday's laws and can be expressed as follows:
\begin{gather}
\hat{\vec{D}} = (c/\omega)^2(-\del\del + \mathbb{1}_3\del^2) + \mathbb{1}_3 + \hat{\vec{\chi}}.
\end{gather}
Here, $c$ is the speed of light, $\omega$ is the wave frequency, $\mathbb{1}_N$ denotes a $N \times N$ unit matrix, and $\hat{\vec{\chi}}$ is the medium susceptibility, which is generally an operator. Let us assume that the characteristic wavelength $\lambda$ of $\vec{E}$ is sufficiently small as characterized by the GO parameter
\begin{gather}
 \bar{\epsilon} \doteq \lambda/\ell \ll 1.
\end{gather}
(Here, the symbol $\doteq$ denotes definitions, and $\ell$ is the smallest length scale of the inhomogeneities, including the wave-envelope length scales.) Under this assumption, \Eq{eq:weq} can be simplified as follows. 

Let us consider the phase-space representation of the dispersion operator. Specifically, we apply the Weyl transform, $\smash{\hat{\vec{D}}} \mapsto \vec{D}$, where the Weyl image $\vec{D}$ is a $(3 \times 3)$-matrix function of the spatial coordinate $\vec{x}$ and the momentum (wave-vector) coordinate $\vec{p}$ \cite{foot:weyl}. Then,
\begin{gather}
\vec{D}(\vec{x}, \vec{p}) = \vec{D}_0(\vec{p}) + \vec{\chi}(\vec{x}, \vec{p}).
\end{gather}
(The dependence on $\omega$ is assumed.) Here, $\vec{D}_0$ corresponds to the vacuum part of the dispersion operator. In terms of components,
\begin{gather}
[\vec{D}_0(\vec{p})]^\mu{}_\nu= (c/\omega)^2(p^\mu p_\nu - p^2\delta^\mu_\nu) + \delta^\mu_\nu,\label{eq:d0}
\end{gather}
where $\mu, \nu = 1,2,3$. (In the Euclidean metric, which is henceforth assumed, upper and lower indexes are interchangeable.) Likewise, $\vec{\chi}$ is the Weyl image of $\hat{\vec{\chi}}$. For clarity, we assume that there is no dissipation, so $\vec{\chi}$ is Hermitian. (Weak dissipation can be introduced additively and does not affect our general approach.) Then, the matrix $\vec{D}(\vec{x}, \vec{p})$ is Hermitian too. From the spectral theorem, it has three orthonormal eigenvectors $\vec{h}_r(\vec{x}, \vec{p})$, and the corresponding eigenvalues $\mc{E}_r(\vec{x}, \vec{p})$ are real, which will be used below.

Let us express the electric field as $\vec{E} = e^{i\theta}\vec{\Psi}$, where $\vec{\Psi}$ is the slow envelope and $\theta$ is the rapid phase. We treat the latter as a prescribed function that remains to be specified (see below). The phase $\theta$ also determines the wave vector $\vec{k} \doteq \del \theta$. By Taylor-expanding $\vec{D}(\vec{x}, \vec{p})$ around $\vec{p} = \vec{k}(\vec{x})$ to the first order in $\bar{\epsilon}$, we get
\begin{gather}\label{eq:Dapprox}
\vec{D}(\vec{x}, \vec{p}) \approx \vec{D}(\vec{x}) + [p_\mu - k_\mu(\vec{x})] \vec{\mc{V}}^\mu(\vec{x}).
\end{gather}
Here, summation over the repeating index $\mu$ is assumed. Also, $\vec{D}(\vec{x}) \doteq \vec{D}(\vec{x}, \vec{k}(\vec{x}))$, $\vec{\mc{V}}^\mu(\vec{x}) \doteq \vec{\mc{V}}^\mu(\vec{x}, \vec{k}(\vec{x}))$, and $\vec{\mc{V}}^\mu(\vec{x}, \vec{p}) \doteq \pd \vec{D}(\vec{x}, \vec{p})/\pd p_\mu$. Then, by applying the inverse Weyl transform to \Eq{eq:Dapprox} and substituting the result in \Eq{eq:weq}, we obtain an approximate envelope equation \cite{foot:part12}
\begin{gather}\label{eq:Dpsi}
\vec{D}(\vec{x})\ccdot\vec{\Psi} - \frac{i}{2}\,
\big[ \pd_\mu \circ \vec{\mc{V}}^\mu(\vec{x}) + \vec{\mc{V}}^\mu(\vec{x})  \circ \pd_\mu
\big]\ccdot\vec{\Psi} \approx 0.
\end{gather}
Here, $\vec{\Psi}$ is a shortened notation for $\vec{\Psi}(\vec{x})$, $\pd_\mu$ is a shortened notation for $\pd/\pd x^\mu$, and the symbol $\circ$ denotes that the operators are applied sequentially. Specifically, $[\pd_\mu \circ \vec{\mc{V}}^\mu(\vec{x})] \ccdot \vec{\Psi}$ means, by definition, that $\vec{\Psi}$ first gets multiplied by $\vec{\mc{V}}^\mu(\vec{x})$ and then the \textit{whole product} is differentiated, so the result is $\pd_\mu [\vec{\mc{V}}^\mu(\vec{x}) \ccdot \vec{\Psi}]$. Likewise, $[\vec{\mc{V}}^\mu(\vec{x}) \circ \pd_\mu \big]\ccdot\vec{\Psi}$ means that $\vec{\Psi}$ is first differentiated and then is multiplied by $\vec{\mc{V}}^\mu(\vec{x})$, so the result is $\vec{\mc{V}}^\mu(\vec{x})(\pd_\mu \vec{\Psi})$. Hence, the two terms differ by $\pd_\mu[\vec{\mc{V}}^\mu(\vec{x})] \ccdot \vec{\Psi}$. Similar expansions were also used, \eg in \Ref{ref:friedland87}.

Since $\vec{\Psi}$ is considered a slow function, the second term in \Eq{eq:Dpsi} is $O(\bar{\epsilon})$ so one can see that
\begin{gather}\label{eq:deps}
\vec{D}(\vec{x})\ccdot\vec{\Psi} = O(\bar{\epsilon}).
\end{gather}
Then, it is convenient to introduce the representation of $\vec{\Psi}$ in the basis formed by the eigenvectors of $\vec{D}(\vec{x})$, namely, $\vec{h}_r(\vec{x}) \doteq \vec{h}_r(\vec{x}, \vec{k}(\vec{x}))$:
\begin{gather}\label{eq:psidec}
\vec{\Psi} = \sum_{r=1}^3 \vec{h}_r(\vec{x}) \psi_r(\vec{x}).
\end{gather}
Here, $\psi_r$ are scalar functions and can be understood as follows. Consider multiplying \Eq{eq:deps} by $\vec{h}_r^\dag(\vec{x})$ from the left. That gives $\mc{E}_r(\vec{x})\psi_r(\vec{x}) = O(\bar{\epsilon})$, where $r = 1,2,3$ (no summation over $r$ is assumed), and $\mc{E}_r(\vec{x}) \doteq \mc{E}_r(\vec{x}, \vec{k}(\vec{x}))$. This shows that, for a given $r$, there are two possibilities: (i)~$\psi_r(\vec{x})$ is small or (ii)~$\mc{E}_r(\vec{x})$ is small. In case (i), the polarization $\vec{h}_r$ does not correspond to a propagating wave mode \textit{per~se}; the small nonzero projection of $\vec{\Psi}$ on $\vec{h}_r$ is only due to the fact that the wave field is not strictly sinusoidal. We call such ``modes'' passive. In case (ii), $(\omega, \vec{k})$ are actually close to those of a wave eigenmode that would exist in a homogeneous medium ($\epsilon = 0$) with the same parameters. In this case, $\psi_r$ can be understood as the local scalar amplitude of $r$th mode of the homogeneous medium, so that $\psi_r = O(1)$ is allowed.

Below, we shall consider two active modes, so the third mode ($r = 3$) is automatically passive. In other words, $\psi_{1,2} = O(1)$ and $\psi_3 = O(\bar{\epsilon})$. [Even if an active mode has a zero amplitude initially, it can be excited later through MC from the other active mode, so its amplitude is considered an $O(1)$ quantity.] With this in mind, let us substitute \Eq{eq:psidec} into \Eq{eq:Dpsi} and multiply the resulting equation from the left by $\vec{\Pi} \doteq \smash{\vec{h}_1\vec{h}_1^\dag} + \smash{\vec{h}_2\vec{h}_2^\dag}$, which is an operator projecting a given vector on the active-mode subspace. Then, one obtains
\begin{gather}\label{eq:aa}
\left[\vec{\mc{E}} - \,i\vec{v} \cdot \nabla  - (i/2)\,(\del \cdot \vec{v}) - \vec{U}\right]\ccdot\vec{\psi} 
\approx 0
\end{gather}
up to a term $O(\bar{\epsilon}^2)$, which is neglected \cite{foot:part12}. Here, 
\begin{gather}\label{eq:adef}
\vec{\psi} \doteq \left(
\begin{array}{c}
\psi_1 \\
\psi_2
\end{array}
\right),
\end{gather}
and we introduced the following $2 \times 2$ diagonal matrices:
\begin{gather}
\vec{\mc{E}}(\vec{x}) \doteq \text{diag}\,\{\mc{E}_1(\vec{x}), \mc{E}_2(\vec{x})\},\label{eq:mcE}\\
\vec{v}(\vec{x}) \doteq  \text{diag}\,\{
\pd_{\vec{p}}\mc{E}_1(\vec{x}, \vec{p}),
\pd_{\vec{p}}\mc{E}_2(\vec{x}, \vec{p})
\}_{\vec{p} = \vec{k}(\vec{x})}.
\end{gather}
Also, $\vec{U}(\vec{x}) = O(\bar{\epsilon})$ is $2 \times 2$ Hermitian matrix given~by
\begin{gather}\label{eq:U}
\vec{U} \doteq - (\vec{\Xi}^{\dag}\ccdot\vec{\mc{V}}^{\mu}\ccdot\pd_{\mu}\vec{\Xi})_A.
\end{gather}
Here, $\vec{\Xi}(\vec{x}) = (\vec{h}_1(\vec{x}), \vec{h}_2(\vec{x}))$ is a non-square, $3 \times 2$ matrix whose columns are $\vec{h}_1$ and $\vec{h}_2$ and the index $A$ stands for ``the anti-Hermitian part of''. Notice that, within the adopted accuracy, $\psi_3$ does not appear in \Eq{eq:aa}, even though $\psi_3$ is generally nonzero. This occurs because the eigenvectors $\vec{h}_r$ are orthogonal to each other.

Equation \eq{eq:aa} can be reduced to an ODE in two limits. One is the limit of a weakly anisotropic medium, which is discussed below. The other one is the limit of a plane wave propagating in a plane-layered medium, possibly at an angle to the inhomogeneity axis. The latter case will not considered here but could be approached similarly.

\subsection{Weakly anisotropic medium}
\label{sec:weak}

Suppose that a medium is only weakly anisotropic, namely, such that $\vec{\mc{E}} = \mc{E}_0\mathbb{1}_2 + O(\bar{\epsilon})$, where $\mc{E}_0$ is a scalar. Then, we can approximate $\vec{v}$ with a scalar matrix, $\vec{v} \approx \vec{v}_0 \mathbb{1}_2$, where $\vec{v}_0 = \smash{[\pd \mc{E}_0 (\vec{x}, \vec{p})/\pd \vec{p}]_{\vec{p} = \vec{k}(\vec{x})}}$ is a vector. [Since $\del \vec{\psi} = O(\bar{\epsilon})$, this approximation introduces an $O(\bar{\epsilon}^2)$ error in \Eq{eq:aa}, but that is beyond the adopted $O(\bar{\epsilon})$ accuracy of our theory.] Then, $\vec{v} \cdot \del \approx \mathbb{1}_2(-2d/dl)$, where $d/dl$ is the appropriately normalized spatial derivative along $\vec{v}_0$, and the factor $-2$ is introduced for convenience (see \Sec{sec:plasma}). Hence, \Eq{eq:aa} becomes an ODE,
\begin{gather}
i\vec{\psi}' = \vec{\trH}\ccdot\vec{\psi}, \label{eq:aeq0} \\
\vec{\trH} \doteq \frac{1}{2}\left[\vec{U} - \vec{\mc{E}} - \frac{i}{2}(\del \cdot \vec{v}_0) \mathbb{1}_2\right],
\end{gather}
where the prime denotes $d/dl$. Let us also split $\vec{\trH}$ into its traceless part $\vec{\mc{H}}$ and the remaining scalar part $\trH_0\mathbb{1}_2$,
\begin{gather}\label{eq:H}
\vec{\mc{H}} = \vec{\trH} - \trH_0\mathbb{1}_2, \quad \trH_0 = \text{Tr}\, \vec{\trH}/2.
\end{gather}
Then, by using the variable transformation
\begin{gather}\label{eq:vt}
\vec{\psi} = \exp\left(-i\int\trH_0\,dl\right)\vec{a},
\end{gather}
one can also represent \Eq{eq:aeq0} as
\begin{gather}\label{eq:aeq}
i \vec{a}' = \vec{\mc{H}}\ccdot\vec{a},
\end{gather}
where the coupling matrix $\vec{\mc{H}}$ can be understood as the wave Hamiltonian. As a reminder, this theory describes the field \textit{in the basis formed by} $\vec{h}_r$ [\Eq{eq:psidec}] rather than in the basis determined by the scalar part of $\vec{D}$ as in the Budden-Kravtsov theory \cite{ref:zheleznyakov83, ref:kravtsov96}.

Since $\vec{\mc{H}}$ is Hermitian, one can readily notice the following. First, \Eq{eq:aeq} has a corollary
\begin{gather}\label{eq:act}
|a_1|^2 + |a_2|^2 = \text{\rm const},
\end{gather}
which reflects the conservation of wave quanta. (More specifically, the conserved quantity is the density of the wave action flux, or equivalently, the energy flux density.) Second, $\vec{\mc{H}}$ can be parameterized as follows:
\begin{gather}
\vec{\mc{H}} = \left(
\begin{array}{cc}
-\alpha & -i\beta\\
i\beta^* & \alpha
\end{array}
\right),
\end{gather}
where $\alpha$ is real. Let us introduce the new variable
\begin{gather}
\vec{q}
\doteq
\left(
\begin{array}{cc}
e^{- i\gamma/2} & 0\\
0 & e^{i\gamma/2}\\
\end{array}
\right)\ccdot
\vec{a},
\end{gather}
where $\gamma \doteq \text{arg}\,\beta$. Let us also replace the independent variable $l$ with ${\tau \doteq \int |\beta|\,dl}$, henceforth called ``time'' for brevity. (Replacing $l$ with $\tau$ can only be done if $\beta$ is nonzero in the whole region of interest.) Then, \Eq{eq:aeq} becomes
\begin{gather}\label{eq:aan2}
i\,\frac{d}{d\tau}\left(
\begin{array}{c}
q_1\\
q_2
\end{array}
\right)
=
\left(
\begin{array}{cc}
-s(\tau) & -i\\
i & s(\tau)
\end{array}
\right)
\left(
\begin{array}{c}
q_1\\
q_2
\end{array}
\right),
\end{gather}
where the dot denotes a derivative with respect to $\tau$ and
\begin{gather}\label{eq:s}
 s \doteq \frac{\alpha}{|\beta|}- \frac{\dot{\gamma}}{2}.
\end{gather} 
Note that the medium parameters enter \Eq{eq:aan2} through only one real function, $s$. Also note that \Eq{eq:aan2} can be equivalently written as a second-order ODE for just one of the mode amplitudes, for example, $q_1$:
\begin{gather}\label{eq:qga}
\ddot{q}_1 + \left(1 + s^2 - i\dot{s}\right)q_1 = 0.
\end{gather}

\subsection{Landau--Zener model}

Equations similar to \Eq{eq:qga} emerge also in other MC theories \cite{book:tracy, ref:tracy03, ref:tracy93, ref:friedland87b, ref:friedland87}. In those works, it is assumed that: (i)~the wave trajectory starts and ends in regions with $s \gg 1$, and, (ii)~in the resonance region ($|s| \lesssim 1$), $s$ can be approximated by a linear function of $\tau$, so \Eq{eq:qga} becomes the Weber equation. This approximation is justified when the time needed for a wave to traverse the resonance region, which is of order $\dot{s}^{-1}$, is small compared to the characteristic time scale of $s$, which is of order one in our units; thus, generally speaking, $\dot{s}$ must be large \cite{foot:lz}. Under these assumptions, the MC problem becomes identical to the Landau--Zener problem \cite{ref:landau32, ref:zener32}. Then, a general asymptotic solution for $\vec{q}$ exists:
\begin{gather}\label{eq:lz1}
\left(
\begin{array}{c}
q_1(+\infty) \\ q_2(+\infty)
\end{array}
\right)
=
\left(
\begin{array}{cc}
\mc{T} & -\mc{C}^*\\
\mc{C} & \mc{T}\\
\end{array}
\right)
\left(
\begin{array}{c}
q_1(-\infty) \\ q_2(-\infty)
\end{array}
\right).
\end{gather}
Here, $|\mc{T}|^2 + |\mc{C}|^2 = 1$, or more specifically,
\begin{gather}\label{eq:lz2}
\mc{T} = \exp(- \pi |\kappa|^2),
\quad
\mc{C} = - \frac{\sqrt{2\pi \kappa}}{\kappa\Gamma(-i|\kappa|^2)},
\end{gather}
where $\Gamma$ is the gamma function, and $\kappa \doteq -i(2\dot{s})^{-1/2}$. 

But note that the Landau--Zener model is not entirely universal. For example, when a wave starts in vacuum, the initial wave numbers of the two modes are identical, so the modes are in resonance from the very beginning. This means that $|s| \lesssim 1$ initially. Likewise, the initial $\dot{s}$ can be small, and then it evolves gradually as the wave enters the medium, so the assumption of constant $\dot{s}$ is inapplicable. Hence, in order to describe MC near boundaries, the Landau--Zener model must be replaced with a different approach, which will be discussed below.

\subsection{Spin analogy}
\label{eq:spin}

Note that \Eq{eq:aan2} is identical to the generic equation describing a two-level quantum system. For example, one can interpret $q_{1,2}$ as the two components of the wave function describing a spin-$1/2$ electron. Then, $S_\mu \doteq \vec{q}^\dag \ccdot \vec{\sigma}_\mu \ccdot \vec{q}$, where $\vec{\sigma}_\mu$ are the Pauli matrices,
\begin{gather}\notag
\vec{\sigma}_x = \left(
\begin{array}{rr}
0 & 1\\
1 & 0
\end{array}
\right)
,\quad
\vec{\sigma}_y = \left(
\begin{array}{rr}
0 & -i\\
i & 0
\end{array}
\right)
,\quad
\vec{\sigma}_z = \left(
\begin{array}{rr}
1 & 0\\
0 & -1
\end{array}
\right),
\end{gather}
serves as the expectation value of the $\mu$th component of the spin vector in units $\hbar/2$. The resulting vector
\begin{gather}
\vec{S} =
\left(
\begin{array}{c}
S_x\\
S_y\\
S_z
\end{array}
\right)
\end{gather}
is akin to but different from the commonly known Stokes vector \cite{ref:kravtsov07, book:born}, since $q_r$ are mode amplitudes rather than electric-field components \textit{per~se}. Importantly, by assuming the parameterization
\begin{gather}
\vec{q} = \bar{q}e^{i\Gamma}
\left(
\begin{array}{c}
e^{-i\upsilon/2}\cos(\varphi/2)\\
e^{i\upsilon/2}\sin(\varphi/2) 
\end{array}
\right),
\end{gather}
where $\smash{\bar{q} \doteq \sqrt{|q_1|^2 +|q_2|^2}}$, $\Gamma \doteq (\text{arg}\,q_1 + \text{arg}\,q_2)/2$, and $\upsilon \doteq \text{arg}\,q_2 - \text{arg}\,q_1$, one has
\begin{gather}
\vec{S} = \bar{q}
\left(
\begin{array}{c}
\sin\varphi\,\cos\upsilon\\
\sin\varphi\,\sin\upsilon\\
\cos{\varphi} 
\end{array}
\right).
\end{gather}
Thus, $|\vec{S}| = \bar{q}$, and one can infer the values of $q_{1,2}$ from a given $\vec{S}$ up to their common phase $\Gamma$.

Now that we have introduced Pauli matrices $\vec{\sigma}_\mu$, let us also use them as a basis to represent the dimensionless Hamiltonian in \Eq{eq:aan2} as follows:
\begin{gather}
\left(
\begin{array}{cc}
-s & -i\\
i & s
\end{array}
\right) = \frac{1}{2}\,\vec{\sigma}_\mu W^\mu.
\end{gather}
The expansion coefficients $W^\mu$ form a three-dimensional vector $\vec{W}$ given by
\begin{gather}
\vec{W} =
\left(
\begin{array}{c}
0\\
2\\
-2s
\end{array}
\right).
\end{gather}
Then, from \Eq{eq:aan2}, one can deduce that $\vec{S}$ is governed by a precession equation \cite{my:qdirac},
\begin{gather}\label{eq:S}
\dot{\vec{S}} = \vec{W} \times \vec{S}.
\end{gather}

Let us assume $\dot{s} \ll 1$. In this case, there is a well-defined local frequency of the complex amplitudes [see \Eq{eq:qga}], which also serves as half of the local precession frequency $\smash{W = 2\sqrt{1+s^2}}$. Then, the orientation of $\vec{S}$, or, more precisely, of the precession plane, follows that of $\vec{W}$ \cite{tex:barnes12}. This is due to the conservation of the adiabatic invariant associated with the precession, so $\dot{s}$ can be viewed as the adiabaticity parameter. Using it, one can construct a Wentzel--Kramers--Brillouin (WKB) asymptotic solution of the MC problem, as shown in the Appendix. Below, we show how it can also be understood without detailed calculations, at least in some limits.

\msection{Nonresonant interaction} Suppose that $s \gg 1$ at all times. Then, $\vec{W}$ remains approximately parallel to the $z$ axis. This means that the precession occurs in the $(x, y)$ plane. If the initial state is a pure mode, \ie the initial $\vec{S}$ is parallel to the $z$ axis, then $\vec{S}$ remains approximately parallel to the $z$ axis (\ie the precession trajectory radius remains zero) forever, so no significant MC is possible. This is understood as a nonresonant interaction. 

\begin{figure}
\centering
\includegraphics[width=.3\textwidth]{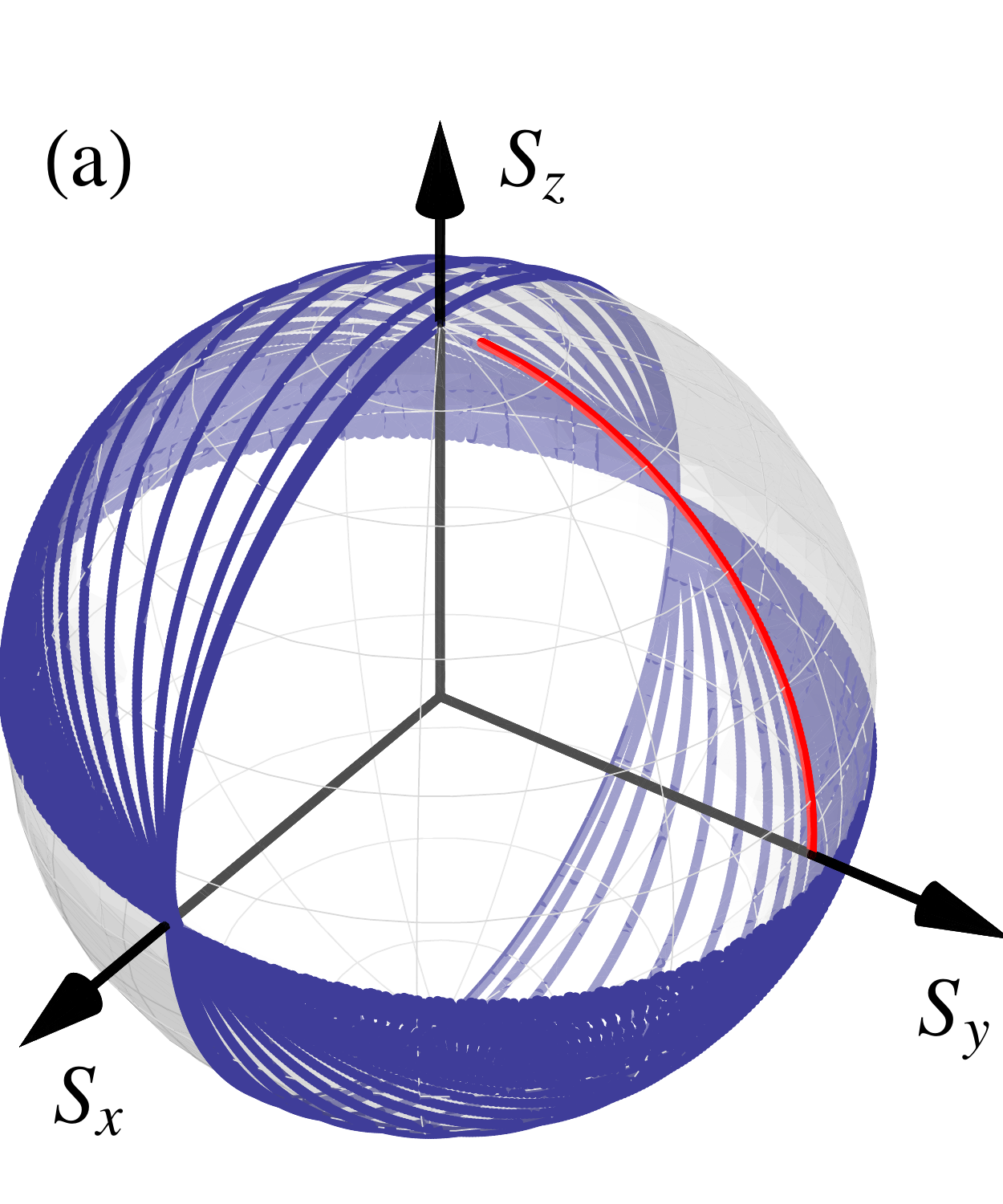}\\
\includegraphics[width=.3\textwidth]{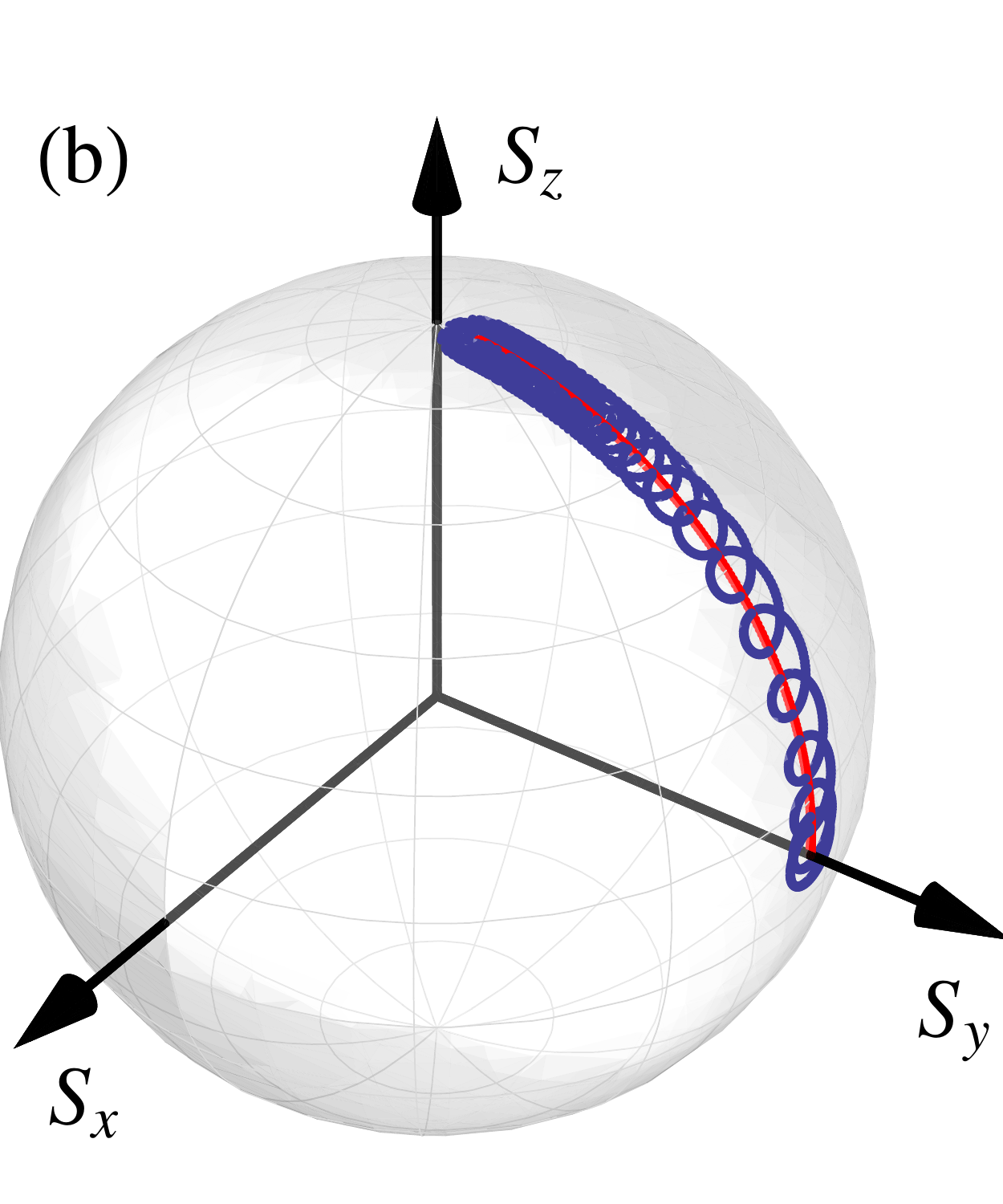}
\caption{Examples of the dynamics of $\vec{S}$ governed by \Eq{eq:S} with a prescribed slowly-changing $\vec{W}$. Shown are the trajectories of $\vec{S}$ (blue) and $\vec{W}/W$ (red) on a unit sphere. (a) Initially, $\vec{S}$ is in the $z$ direction and $\vec{W}$ is in the $y$ direction. Later, $\vec{W}$ rotates and becomes pointed in the $z$ direction. This causes the precession plane to rotate accordingly, so eventually, the vector $\vec{S}$ ends up precessing in the horizontal plane. Hence, the final state is a mixed-mode state. (b) Both $\vec{S}$ and $\vec{W}$ are in the $y$ direction initially and then slowly rotate together towards the $z$ axis. The final state corresponds to $S_z = 1$, which is a pure-mode state.}
\label{fig:s0}
\end{figure}

\msection{Resonant interaction} Now suppose that a wave starts at $s = 0$, so the initial $\vec{W}$ is in the $y$ direction. As the wave enters a medium, $s$ begins to grow and $\vec{W}$ starts rotating in the $(y, z)$ plane. Suppose that the dispersion curves of the two modes grow apart eventually, so $s$ becomes large (${s \gg 1}$). Then, the final direction of $\vec{W}$ is along the $z$ axis. If the initial state is a pure mode, \ie the initial $\vec{S}$ is parallel to the $z$ axis, then $\vec{S}$ starts precessing in the $(x, z)$ plane first, but eventually the precession plane orients transversely to the $z$ axis [\Fig{fig:s0}(a)]. This means that, in the final state, the two modes have equal amplitudes. Conversely, in order to obtain a pure state when $s \gg 1$, one needs to start with a mixed-mode state corresponding to $\vec{S}$ along the $y$ axis. As shown in \Fig{fig:s0}(b), as the vector $\vec{W}$ starts rotating in the $(y, z)$ plane, the vector $\vec{S}$ precesses around $\vec{W}$. The final state of $\vec{S}$ is aligned to the $z$ axis, which is a pure state mode. 

These general arguments can also be reformulated in terms of our original variables $\vec{a}$ instead of $\vec{q}$. In this case, we define
\begin{gather}
 S_\mu \doteq \vec{a}^\dag \ccdot \vec{\sigma}_\mu \ccdot \vec{a}
\end{gather}
and use \Eq{eq:aeq} to get
\begin{gather}
\vec{S}' = \vec{W} \times \vec{S}
\end{gather}
(note that the independent variable here is $l$ rather than $\tau$; hence the prime), where the new vector $\vec{W}$ is
\begin{gather}\label{eq:W}
\vec{W} =
2\left(
\begin{array}{c}
\text{Im}\,\beta\\
\text{Re}\,\beta\\
-\alpha
\end{array}
\right).
\end{gather}
The vacuum value of $\vec{W}$ is not necessarily along the $y$ axis now. Still, if $\vec{W}$ evolves slowly ($\dot{s} \ll 1$), the method for ensuring a single-mode operation in the large-$s$ region is the same as before, namely, $\vec{S}$ must be initialized parallel to $\vec{W}$. 

\section{Low-density plasma}
\label{sec:plasma}

In this section, we apply the above general theory to EM waves in cold magnetized low-density plasma.

\subsection{Basic approximations}

Suppose a low-density plasma such that $\vec{\chi}$ is comparable to $\bar{\epsilon}$. In this case, we can choose $\theta$ such that $\vec{k}$ is constant and satisfies the vacuum dispersion relation
\begin{gather}\label{eq:vack}
k^2 = \omega^2/c^2.
\end{gather}
(This does \textit{not} mean that the plasma dispersion is neglected. We simply choose to describe it as an effect on $\vec{a}$ rather than as an effect on $\theta$. As long as $\vec{a}$ remains slow, these descriptions are equivalent.) We also choose $\vec{D}_0$ for $\mc{E}_0\mathbb{1}_2$ (\Sec{sec:weak}). Then, $\mc{E}_0(\vec{x}, \vec{p}) = 1 - (pc/\omega)^2$, $\vec{v}_0 =- 2 \vec{k}/k^2$, $\del \cdot \vec{v}_0 = 0$, and
\begin{gather}
\frac{d}{dl} = \frac{k^\mu}{k^2}\, \frac{\pd}{\pd x^\mu} = \frac{\vec{e}_k \cdot \del}{k},
\quad
\vec{e}_k \doteq \frac{\vec{k}}{k}.
\end{gather}

The true eigenvalues of $\vec{D}(\vec{x}, \vec{p})$ in the plasma can be found by considering $\vec{\chi}$ as a small perturbation to $\vec{D}_0$ and by using the standard perturbation theory \cite{book:landau3}:
\begin{gather}
\mc{E}_r(\vec{x}, \vec{p}) \approx \mc{E}_0(\vec{x}, \vec{p}) + 
(\vec{\eta}_r^\dag \ccdot \vec{\chi} \ccdot \vec{\eta}_r)(\vec{x}, \vec{p}).
\end{gather}
Here, $\vec{\eta}_r$ is the zero-density limit of $\vec{h}_r$, and $r = 1,2$ (henceforth assumed). Since ${\mc{E}_0(\vec{x}, \vec{k}) = 0}$, this gives 
\begin{gather}\label{eq:Er}
\mc{E}_r(\vec{x}) = (\vec{\eta}_r^\dag \ccdot \vec{\chi} \ccdot \vec{\eta}_r)(\vec{x}, \vec{k}(\vec{x})).
\end{gather}

In order to calculate $\vec{U}$, which is already of the first order in $\bar{\epsilon}$, let us use the zeroth-order approximation for $\vec{\mc{V}}^\mu$, namely, $\vec{\mc{V}}^\mu(\vec{x}, \vec{p}) \approx \pd \vec{D}_0/\pd p_\mu$ \cite{foot:coldV}. This gives
\begin{gather}
\vec{\mc{V}}^\mu = \frac{c^2}{\omega^2}\,[
(\vec{e}^\mu)\vec{k}^\dag + \vec{k} (\vec{e}^\mu)^\dag - 2k^\mu \mathbb{1}_3
],
\label{eq:V}
\end{gather}
where $\vec{e}^\mu$ is a unit vector along the $\mu$ axis. [The adjoints of the real vectors $\vec{e}^\mu$ and $\vec{k}$ are the row vectors obtained simply by transposing $\vec{e}^\mu$ and $\vec{k}$, correspondingly. Hence, for example, $\smash{(\vec{e}^\mu)\vec{k}^\dag}$ is a $3 \times 3$ matrix but $\smash{\vec{k}^\dag(\vec{e}^\mu)}$ is a scalar.] Let us also use the approximation $\vec{\Xi} \approx (\vec{\eta}_1, \vec{\eta}_2)$. Since $\vec{\eta}_1$ and $\vec{\eta}_2$ are \textit{vacuum} polarization vectors, they are orthogonal to $\vec{k}$,~so
\begin{gather}
\vec{k}^\dag \ccdot \vec{\Xi} 
= (\vec{k}^\dag \ccdot\vec{\eta}_1, \vec{k}^\dag \ccdot \vec{\eta}_2)
= 0
\end{gather}
and, similarly, $\vec{\Xi}^\dag  \ccdot \vec{k} = 0$. From this and the fact that $\vec{k}$ is constant, it is readily seen that the first two terms in \Eq{eq:V} do not contribute to $\vec{U}$ in \Eq{eq:U}. This leads~to
\begin{gather}\label{eq:U2}
\vec{U} = 2(\vec{\Xi}^{\dag} \ccdot \vec{\Xi}')_A,
\end{gather}
where we invoked \Eq{eq:vack}. Then, using the fact that
\begin{gather}\notag
\vec{\eta}_1^\dag \ccdot \vec{\eta}_1
= \vec{\eta}_2^\dag \ccdot \vec{\eta}_2 
= 1,\quad
\vec{\eta}_1^\dag \ccdot \vec{\eta}_2
= \vec{\eta}_2^\dag \ccdot \vec{\eta}_1
= 0,
\end{gather}
one can express $\vec{U}$ as follows:
\begin{gather}\label{eq:U3}
\vec{U} = 
-2i
\left(
\begin{array}{cc}
\vec{\eta}_1^\dag \ccdot \vec{\eta}_1' 
& \vec{\eta}_1^\dag \ccdot \vec{\eta}_2'
\\
- (\vec{\eta}_1^\dag \ccdot \vec{\eta}_2')^*
& \vec{\eta}_2^\dag \ccdot \vec{\eta}_2'
\end{array}
\right).
\end{gather}

\subsection{Polarization vectors $\boldsymbol{\vec{\eta}_{1,2}}$}
\label{sec:etas}

Let us now explicitly calculate $\vec{\eta}_{1,2}$ assuming the plasma is cold. In order to do so, let us temporarily adopt coordinates such that the local dc magnetic field $\vec{B}_0$ is along the $z$~axis, and the $y$~axis is orthogonal to the plane formed by $\vec{B}_0$ and $\vec{k}$; \ie
\begin{gather}
\vec{b} = \left(
\begin{array}{c}
0\\
0\\
1
\end{array}
\right),
\quad
\vec{k} = \left(
\begin{array}{c}
k \sin \vartheta\\
0\\
k \cos \vartheta
\end{array}
\right).
\end{gather}
(The general result for arbitrary direction of $\vec{k}$ and $\vec{B}_0$ is presented at the end of this section.) Here, $\vec{b} \doteq \vec{B}_0/B_0$ is a unit vector along $\vec{B}_0$, and $\vartheta$ is the angle between $\vec{k}$ and $\vec{B}_0$. Then \cite{book:stix},
\begin{gather}\label{eq:chi}
\vec{\chi} = \left(
\begin{array}{ccc}
S - 1 & -iD & 0\\
iD & S - 1  & 0\\
0 & 0 & P - 1
\end{array}
\right).
\end{gather}
We shall limit our consideration to high-frequency waves, so the ion response can be neglected entirely. Hence \cite{book:stix},  
\begin{gather}\notag
S = 1 - \frac{\omega_p^2}{\omega^2 - \Omega^2},\quad
D = \frac{\Omega}{\omega}\,\frac{\omega_p^2}{\omega^2 - \Omega^2},\quad
P = 1 - \frac{\omega_p^2}{\omega^2},
\end{gather}
where $\omega_p \doteq (4\pi n_e e^2/m)^{1/2}$ is the electron plasma frequency, $\Omega \doteq eB_0/(mc)$ is the electron gyrofrequency, $n_e$ is the electron density, $e < 0$ is the electron charge, and $m$ is the electron mass. Also \cite{book:stix},
\begin{gather}\notag
\vec{D}(\vec{x}, \vec{k}) = \left(
\begin{array}{ccc}
S - N^2\cos^2\vartheta & -iD & N^2\sin\vartheta\cos\vartheta\\
iD & S - N^2  & 0\\
N^2\sin\vartheta\cos\vartheta & 0 & P - N^2\sin^2\vartheta
\end{array}
\right),
\end{gather}
where $N \doteq k\omega/c$ is the refraction index. From $\vec{D}(\vec{x}, \vec{k}) \ccdot \vec{\eta}_{1,2} = 0$, we find that each of $\vec{\eta}_{1,2}$ satisfies
\begin{gather}
\frac{\eta_y}{\eta_x}  = - \frac{iD}{S - N^2},\\
\frac{\eta_z}{\eta_x}  = - \frac{N^2\sin\vartheta\cos\vartheta}{P - N^2\sin^2\vartheta}.
\end{gather}
Also, the dispersion relation $\det\vec{D} = 0$ gives \cite{book:stix}
\begin{gather}
N^2_{1,2} = \frac{B \pm \sqrt{B^2 - 4AC}}{2A},
\end{gather}
where
\begin{gather}
A = S \sin^2\vartheta + P \cos^2\vartheta,\\
B = (S^2 - D^2) \sin^2\vartheta + PS (1 + \cos^2\vartheta),\\ 
C = P(S^2 - D^2).
\end{gather}
Then, in the zero-density limit, one obtains \cite{foot:math}
\begin{gather}\label{eq:etas1}
\vec{\eta}_1 = \frac{1}{\sqrt{1 + g_1^2}}\left(
\begin{array}{c}
-\cos\vartheta \\
ig_1 \\
\sin\vartheta
\end{array}
\right),\\
\vec{\eta}_2 = -\frac{i\varsigma}{\sqrt{1 + g_2^2}}\left(
\begin{array}{c}
-\cos\vartheta \\
ig_2 \\
\sin\vartheta
\end{array}
\right),\label{eq:etas2}
\end{gather}
where we introduced
\begin{gather}
g_{1,2} \doteq u^{-1} \mp \varsigma \sqrt{1 + u^{-2}}, \quad \\
u \doteq \frac{2\omega}{\Omega}\, \csc\vartheta\cot\vartheta,\label{eeq:u}
\end{gather}
and $\varsigma \doteq \text{sgn}\,u$. The normalization in $\vec{\eta}_{1,2}$ is chosen such that, at $\vartheta \to \pi/2$, one gets $\vec{\eta}_1 \to \vec{e}_z$ and $\vec{\eta}_2 \to \vec{e}_y$, which corresponds to the O and X waves, correspondingly. (Here, $\vec{e}_\mu$  denotes a unit vector along the axis $\mu$. Also remember that our axes are tied to the local $\vec{B}_0$.) Accordingly, at $\vartheta \to 0$, one gets $\vec{\eta}_1 \to \smash{-(\vec{e}_x + i\vec{e}_y\, \text{sgn}\, \Omega)/\sqrt{2}}$ and $\vec{\eta}_2 \to \smash{(i\,\text{sgn}\,\Omega)(\vec{e}_x - i\vec{e}_y\,\text{sgn}\,\Omega)/\sqrt{2}}$. Then, $\vec{\eta}_1$ corresponds to the L wave at $\Omega < 0$ (R wave at $\Omega > 0$), and $\vec{\eta}_2$ corresponds to the R wave at $\Omega < 0$ (L wave at $\Omega > 0$).

Finally, let us rewrite $\vec{\eta}_{1,2}$ in the invariant form. To do this, note that $\vec{k} = k_x\vec{e}_x + k_z \vec{e}_z$ and $\vec{e}_z = \vec{b}$, so
\begin{gather}
\vec{e}_x = \frac{\vec{k} - \vec{b} k_z}{k_x} = \frac{\vec{e}_k - \vec{b}\cos\vartheta}{\sin\vartheta}.
\end{gather}
Then, one obtains
\begin{gather} 
\vec{\eta}_1 = \frac{\vec{e}_* + ig_1 \vec{e}_y}{\sqrt{1 + g_1^2}},\quad
\vec{\eta}_2 = -\frac{i\varsigma(\vec{e}_* + ig_2\vec{e}_y)}{\sqrt{1 + g_2^2}},
\end{gather}
where $\vec{e}_*$ is introduced, merely to shorten the notation, as a unit vector given by $\vec{e}_* \doteq (\vec{b} - \vec{e}_k\cos\vartheta)/\sin\vartheta$. Fully invariant expressions for $\vec{\eta}_{1,2}$ can be obtained by using $\vec{e}_y = (\vec{b} \times \vec{e}_k)/\sin\vartheta$ (this equality is seen from the fact that, by definition, $\vec{e}_y$ is orthogonal to both $\vec{b}$ and $\vec{k}$ and has a unit length) and
\begin{gather} \label{eq:cossin}
\cos\vartheta = \vec{e}_k \cdot \vec{b}, \quad 
\sin\vartheta = |\vec{e}_k \times \vec{b}|.
\end{gather}

\subsection{Wave Hamiltonian}
\label{sec:explicit}

Using \Eq{eq:Er} along with \Eqs{eq:etas1} and \eq{eq:etas2}, one readily finds that
\begin{multline}\label{eq:e12}
\mc{E}_{1,2}
= - \frac{\omega_p^2}{\omega^2 - \Omega^2} \bigg[
1 - \frac{\Omega^2}{2\omega^2}\, (\vec{e}_k \times \vec{b})^2
\\
\mp \frac{|\Omega|}{\omega} 
\sqrt{(\vec{e}_k \cdot \vec{b})^2 +\frac{\Omega^2}{4\omega^2} \,(\vec{e}_k \times \vec{b})^4}
\bigg].
\end{multline}
Note that $\mc{E}_{1,2}$ are linear with respect to the plasma density due to the perturbative approach [\Eq{eq:Er}]. By using \Eq{eq:U3}, one also finds that 
\begin{gather}\label{eq:Use}
\vec{U} = 2\left(
\begin{array}{cc}
\epsilon_z & \epsilon_x - i\epsilon_y\\
\epsilon_x + i\epsilon_y & -\epsilon_z
\end{array}
\right),
\end{gather}
where $\epsilon_i = O(\bar{\epsilon})$ are scalar functions given~by
\begin{gather}\label{eq:epsilons}
\epsilon_x = - \frac{u'}{2(1 + u^2)} = \frac{\delta'}{2}, \\
\epsilon_y = \frac{\zeta}{\sqrt{1 + u^2}} = \zeta \cos \delta,\\
\epsilon_z = - \frac{\zeta u}{\sqrt{1 + u^2}} = \zeta \sin \delta.
\end{gather}
Here, $\delta \doteq - \text{arctan}\,u$ [where $u$ is given by \Eq{eeq:u}] and $\zeta \doteq \vec{e}_* \cdot \vec{e}_y'$, or, more explicitly,
\begin{gather}\label{eq:wzeta}
u \doteq \frac{2\omega (\vec{e}_k \cdot \vec{b})}{\Omega (\vec{e}_k \times \vec{b})^2},\quad
\zeta \doteq \frac{(\vec{e}_k \times \vec{b}) \cdot \vec{b}'}{(\vec{e}_k \times \vec{b})^2},
\end{gather}
so $\zeta$ serves as a measure of the magnetic shear. (The coefficient $\epsilon_x$ is the same at the one known for shearless fields, where $\zeta = 0$ \cite{ref:zheleznyakov79}.) Also, $\rho \doteq \mc{E}_1 - \mc{E}_2$, namely,
\begin{gather}
\rho
= \frac{2|\Omega|\omega_p^2}{\omega(\omega^2 - \Omega^2)} 
\sqrt{(\vec{e}_k \cdot \vec{b})^2 +\frac{\Omega^2}{4\omega^2} \,(\vec{e}_k \times \vec{b})^4}.
\end{gather}

Hence, the traceless Hamiltonian $\vec{\mc{H}}$ that governs MC [\Eq{eq:H}] is as follows:
\begin{gather}
\vec{\mc{H}} = \left(
\begin{array}{cc}
\epsilon_z - \rho/2 & \epsilon_x - i\epsilon_y\\
\epsilon_x + i\epsilon_y & -\epsilon_z + \rho/2
\end{array}
\right).
\end{gather}
Accordingly,
\begin{gather}\label{eq:ab}
\alpha = \frac{\rho}{2} - \epsilon_z,
\quad 
\beta = \epsilon_y + i \epsilon_x,
\quad
\gamma = \text{arctan}\left(\frac{\epsilon_x}{\epsilon_y}\right).
\end{gather}
Also, the corresponding vector $\vec{W}$ [\Eq{eq:W}] that governs the precession of $\vec{S}$ is
\begin{gather}
\vec{W} = \left(
\begin{array}{c}
2\epsilon_x \\ 2\epsilon_y \\ 2\epsilon_z - \rho
\end{array}
\right).
\end{gather}

\subsection{Propagation parallel to $\boldsymbol{\vec{B}_0}$}
\label{sec:paral}

When a wave propagates parallel to a dc magnetic field (\ie $\vartheta = 0$), \Eq{eq:e12} gives
\begin{gather}
\mc{E}_{1,2} = - \frac{\omega_p^2}{\omega(\omega \mp \Omega)},
\end{gather}
and \Eqs{eq:Use}-\eq{eq:wzeta} give $\vec{U} = 0$. In this particular case, the two modes are uncoupled, and \Eq{eq:aeq0} leads to
\begin{gather}
\psi_r = \exp\left(\frac{i}{2}\int \mc{E}_r\,dl\right)\psi_{r0},
\end{gather}
where the constants $\psi_{r0}$ are determined by the initial conditions. Hence, each $|\psi_r|$ is conserved and $\mc{E}_r/2$ serves a correction to the refraction index. (As a reminder, $dl$ is a ray-path element measured in units $k^{-1} = c/\omega$.) The total refraction indexes in this case are
\begin{gather}
N_{1,2} \approx 1 - \frac{\omega_p^2}{2\omega(\omega \mp \Omega)}.
\end{gather}
This is in agreement with the low-density asymptotics of the known L- and R-wave refraction indexes~\cite{book:stix}. 

\subsection{Propagation perpendicular to~$\boldsymbol{\vec{B}_0}$}
\label{sec:perp}

When a wave propagates perpendicular to a dc magnetic field (\ie $\vartheta = \pi/2$), \Eq{eq:e12} gives
\begin{gather}
\mc{E}_1 = - \frac{\omega_p^2}{\omega^2},\quad
\mc{E}_2 = - \frac{\omega_p^2}{\omega^2 - \Omega^2}.
\end{gather}
Then, one obtains
\begin{gather}
\rho = \frac{\omega_p^2 \Omega^2}{\omega^2(\omega^2 - \Omega^2)}.
\end{gather}
Also, in this case, one has $u = 0$, so $\epsilon_x = \epsilon_z = 0$, and
\begin{gather}
\epsilon_y = \zeta = (\vec{e}_k \times \vec{b}) \cdot \vec{b}'.
\end{gather}
The corresponding function $s$ that enters \Eq{eq:aan2} is ${s  = \rho/(2|\zeta|)}$, and ${\tau' = |\zeta|}$. The adiabaticity parameter in this case is $\dot{s} \sim \rho'/\zeta^2$, where we assumed constant shear (more specifically, $\zeta'/\zeta \ll \rho'/\rho$). We can estimate $\rho'$ as $\rho/(k L_n)$, where $L_n$ is the characteristic scale of the plasma density profile. Here, $\rho$ can be evaluated at the edge of the adiabaticity domain (${s \sim 1}$). This gives ${\rho \sim \zeta}$. Then, ${\dot{s} \sim L_b/L_n}$, where ${L_b \doteq |k\zeta|^{-1}}$ is the characteristic scale of the magnetic-field shear. If $L_n \lesssim L_b$ (weak magnetic shear), then ${\dot{s} \gtrsim 1}$, so the wave leaves the resonance region before it has time to mode-convert (as long as the GO approximation is satisfied, \ie ${L_n \gg \lambda}$). Then, a wave that is initially a pure mode remains such upon entering dense plasma. In contrast, if ${L_b \ll L_n}$ (strong magnetic shear), then $\dot{s} \ll 1$, so the wave ``spin'' $\vec{S}$ follows $\vec{W}$ (\Sec{eq:spin}). Then, substantial MC is possible. In particular, this explains the results presented in \Ref{ref:kubo15}. 

Let us consider the case of strong magnetic shear. Then, according to the argument in \Sec{eq:spin}, a wave that is a pure mode in vacuum eventually transforms into a mixture of the O and X waves with equal amplitudes ($|a_1| = |a_2|$). Conversely, a wave that is composed of two modes in vacuum can asymptotically transform into a single-mode wave upon entering high-density plasma. In the case of the O~wave, this requires that the initial amplitudes satisfy \Eq{eq:ini2}; \ie the initial vacuum wave must be circularly polarized. Ending up with a pure X~wave instead of a pure O~wave requires starting with the opposite circular polarization. 

For the case of general wave propagation with respect to the magnetic field (arbitrary $\vartheta$), the MC process can be understood similarly except that $s$ is given by a more general formula [\Eqs{eq:s} and \eq{eq:ab}].

\section{Conclusions}
\label{sec:conc}

In summary, we developed a theory of EM mode conversion in cold low-density plasma, specifically, in the high-frequency limit where the ion response is negligible. In contrast to the classic (Landau--Zener-type) theory of mode conversion, the region of resonant coupling in low-density plasma is not necessarily narrow, so the coupling matrix cannot be approximated with its first-order Taylor expansion; also, the initial conditions are set up differently. For the case of strong magnetic shear, a simple method is identified for preparing a two-mode wave such that it transforms into a single-mode wave upon entering a high-density plasma. The theory can be used for reduced modeling of wave-power input in fusion plasmas. In particular, applications are envisioned in stellarator research, where the mutual conversion of two EM modes near the plasma edge is a known issue~\cite{ref:kubo15, ref:notake05, ref:tsujimura15}.

The first author (IYD) acknowledges the support and hospitality of the National Institute of Fusion Science, Japan. This work was also supported by the U.S. DOE through Contract No. DE-AC02-09CH11466 and by the U.S. DOD NDSEG Fellowship through Contract No. 32-CFR-168a.

\appendix

\begin{figure}
\centering
\includegraphics[width=.48\textwidth]{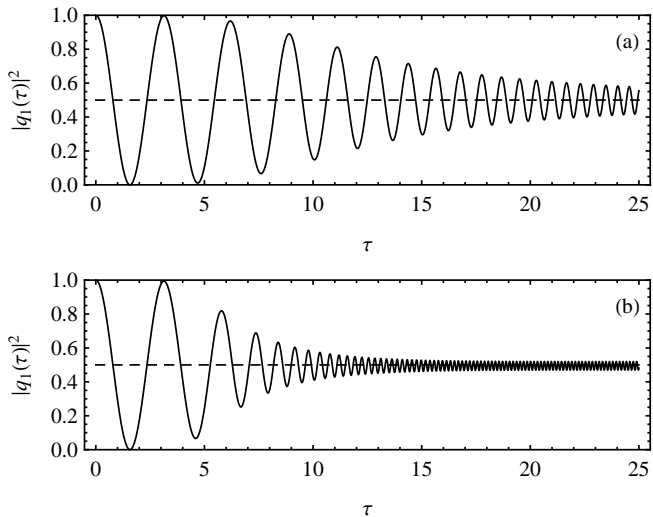}
\caption{Numerical solutions of \Eq{eq:Q} for $|q_1|^2$ for two sample profiles: (a)~$\smash{s(\tau) = 0.02 \tau^2}$ and (b)~$s(\tau) = \smash{40 \tanh[(\tau/15)^3]}$. The envelope $q_1$ is measured in units $\bar{q}$ such that $|\bar{q}|^2$ is the total number of quanta, which is conserved [\Eq{eq:act}]. The initial conditions are $q_1(0) = 1$ and $q_2(0) = 0$, \ie $\dot{q}_1(0) = 0$. Both solutions approach the same asymptote $|q_1|^2 = 1/2$ (dashed), as predicted [\Eq{eq:aov}].}
\label{fig:A}
\end{figure}

\section{WKB model}
\label{app:wkb}

\subsection{Basic equations}

Here, we present a formal WKB derivation of the slow ``spin-precession'' dynamics discussed in \Sec{eq:spin}. We shall refer to the medium as plasma, and $s$ will be treated as a measure of the plasma density. We also adopt that $s(\tau = 0) = 0$ corresponds to vacuum. (This is the case for the example considered in \Sec{sec:perp}.) However, the general idea holds in a broader context too. Also notably, the following model can be understood as a generalization of the ``helical-wave'' GO discussed in \Ref{ref:kocharovskii80}.

Let us start with rewriting \Eq{eq:qga} as
\begin{gather}\label{eq:Q}
\ddot{q}_1 + Q(\tau) q_1 = 0, \quad
Q \doteq 1 + s^2 - i\dot{s}.
\end{gather}
Suppose that
\begin{gather}\label{eq:wkb}
\frac{d}{d\tau}\left(\frac{2\pi}{\sqrt{Q}}\right) \ll 1,
\end{gather}
a sufficient condition for which is $\dot{s} \ll 1$. Then, the WKB approximation is applicable,
\begin{gather}\label{eq:wkb0}
q_1(\tau) \approx  \frac{1}{[Q(\tau)]^{1/4}}\left[
C_+ e^{i\phi(\tau)} + C_-\, e^{-i\phi(\tau)}
\right],
\end{gather}
where $C_{\pm}$ are constants determined by the initial conditions at $\tau = 0$, and $\phi \doteq \smash{\int_0^\tau \sqrt{Q(\tilde{\tau})}\,d\tilde{\tau}}$. Since $\dot{s}$ is assumed small, we Taylor-expand $\sqrt{Q}$ to~get $\phi = \phi_{\rm re} + i \phi_{\rm im}$, where
\begin{gather}
\phi_{\rm re} \approx \sigma_s \int_0^\tau \sqrt{1+s^2(\tilde{\tau})} \,d\tilde{\tau},\\
\phi_{\rm im} = -\sigma_s \int^s_0 \frac{d\tilde{s}}{2\sqrt{1 + \tilde{s}^2}} 
= -\frac{1}{2}\ln\left(|s| + \sqrt{1 + s^2}\right).\label{eq:phii}
\end{gather}
The sign of $\phi$ is a matter of convention and depends on the interpretation of $C_{\pm}$. We introduced a sign factor $\sigma_s \doteq \text{sgn}\,s(\tau)$ only to ensure that, at $s \gg 1$, one can unambiguously identify the term $\propto e^{i\phi}$ as the first mode (Mode~I) and the term $\propto e^{-i\phi}$ as the second mode (Mode~II), as seen from \Eq{eq:aan2}. At $s \lesssim 1$, both terms contribute to both modes.

Within this WKB model, if a wave starts \textit{and} ends outside plasma, the mode amplitude is preserved; namely, $\phi_{\rm im}(\infty) = 0$ due to $s(\infty) = 0$. However, note that this requires the low-density approximation to hold at all $\tau$, which is usually not the case. As a rule, a wave eventually enters a high-density region where rays of the two modes diverge or some dissipation occurs. Thus, even if the radiation escapes plasma later, the original single-mode wave is not quite restored. Hence, the process of rays leaving the plasma will not be considered. 

\subsection{Starting with a pure mode}
\label{sec:pure}

Suppose a wave outside plasma is a pure Mode~I, so $q_1(0) = \bar{q}$, $q_2(0) = 0$, and $s(0) = \dot{s}(0) = 0$, so $\dot{q}_1(0) = 0$ and $Q(0) = 1$. Then, $C_+ = C_- = \bar{q}/2$, so \Eq{eq:wkb0} gives
\begin{gather}
\frac{q_1(\tau)}{\bar{q}} \approx  \frac{\cos \phi(\tau)}{[Q(\tau)]^{1/4}}.
\end{gather}
In the high-density limit ($s \gg 1$), one has $Q \approx s^2$, $2|\cos\phi| \approx \exp(|\phi_{\rm im}|)$, and $2|\phi_{\rm im}| \approx \ln(2|s|)$. Then, $|q_1/\bar{q}|^2$ asymptotically approaches a universal constant:
\begin{gather}\label{eq:aov}
\left|\frac{q_1(\infty)}{\bar{q}}\right|^2 \approx \frac{e^{2|\phi_{\rm im}|}}{4|s|} \approx \frac{1}{2}.
\end{gather}
Due to \Eq{eq:act}, one also gets $\smash{|q_2(\infty)/\bar{q}|^2} \approx 1/2$. Thus, during the MC, an initially-pure Mode~I asymptotically loses half of its action flux (which we also term loosely as ``quanta'') to Mode~II. This conclusion also agrees with numerical calculations (\Fig{fig:A}), and similar results apply if the the initial conditions of the two modes are interchanged.

\subsection{Ending with a pure mode}

Suppose now that a wave \textit{becomes} a pure Mode~I asymptotically in the high-density limit ($s \gg 1$). This means that $C_- = 0$. Also, using \Eq{eq:phii}, we obtain
\begin{gather}
q_1 = \frac{C_+}{Q^{1/4}}\,e^{i\phi} \approx C_+e^{i\phi_{\rm re}}\sqrt{1 + \frac{s}{\sqrt{1+s^2}}}.
\end{gather}
In the limit $s \gg 1$, this gives $q_1 \approx C_+\sqrt{2} e^{i\phi_{\rm re}} = \bar{q} e^{i\phi_{\rm re}}$, where $\bar{q} \doteq C_+\sqrt{2}$ can be interpreted as the asymptotic amplitude of Mode~I deep inside the plasma. Then,
\begin{gather}\label{eq:ini}
q_1(0) \approx \frac{\bar{q}}{\sqrt{2}}, 
\quad
\dot{q}_1(0) \approx \frac{i \sigma_s\bar{q}}{\sqrt{2}}, 
\end{gather}
where we used $Q = 1$ for vacuum [and thus $\dot{\phi}(0) = 1$]. Since $q_2 = -\dot{q}_1 + is q_1$ [\Eq{eq:aan2}], one can also rewrite \Eqs{eq:ini}~as
\begin{gather}\label{eq:ini2}
q_1(0)/q_2(0) \approx i \sigma_s,
\end{gather}
which corresponds to $S_y(0) = \pm 1$. These initial conditions ensure that a wave that is initially a mixture of Modes~I and~II asymptotically converts into the pure Mode-I upon entering high-density plasma.

\begin{figure}
\centering
\includegraphics[width=.48\textwidth]{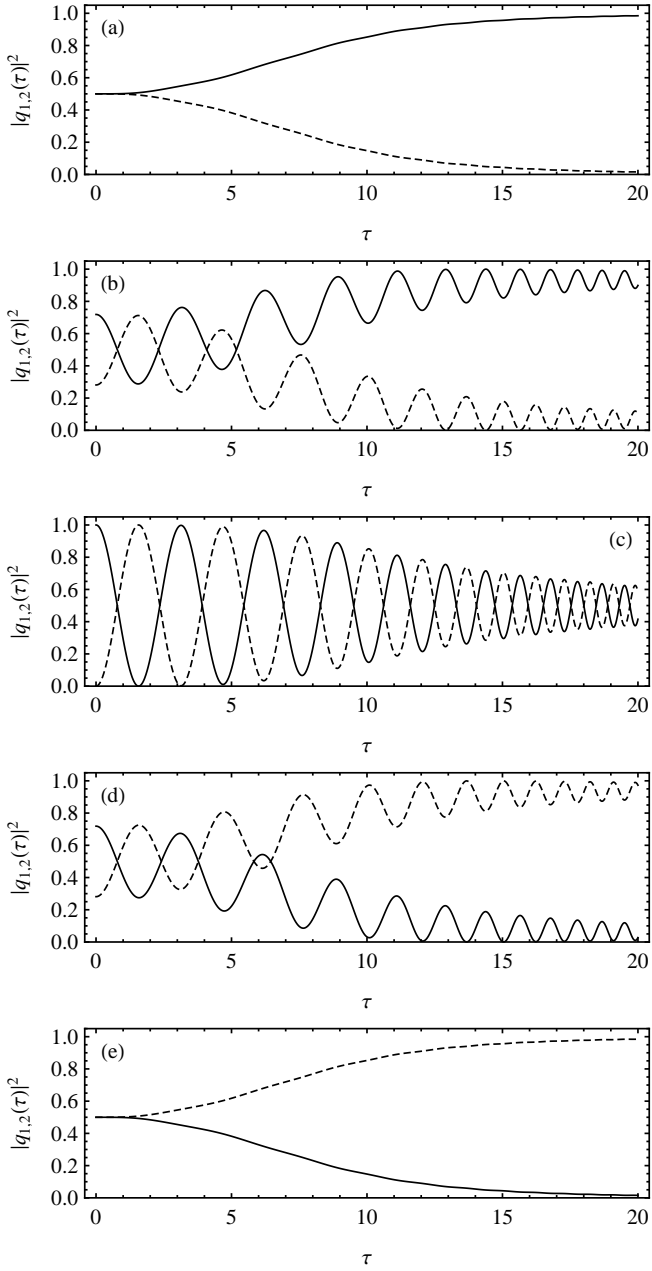}
\caption{Numerical solutions of \Eq{eq:Q} for $|q_1|^2$ (solid) and $|q_2|^2$ (dashed) for $s(\tau) = 0.02 \tau^2$. The envelopes $q_{1,2}$ are measured in units $\bar{q}$ such that $|\bar{q}|^2$ is the total number of quanta, which is conserved [\Eq{eq:act}]. The initial conditions are $q_1(0) = \smash{\sqrt{1 - u^2/2}}$ and $\dot{q}_1 \equiv - q_2(0) = i u/\sqrt{2}$, so $|q_1(0)|^2 + |q_2(0)|^2 = 1$. Here: (a)~${u = 1}$, (b)~${u = 3/4}$, (c)~${u = 0}$, (d)~${u = -3/4}$, and (e)~${u = -1}$. In the cases (a) and (e), one starts with a circularly polarized wave in vacuum and ends up with a single-mode wave in dense plasma. In the case (c), where the solid curve is the same as in \Fig{fig:A}(a), one starts with a pure Mode~I and ends up with a mixture of Modes~I and~II with equal number of quanta.}
\label{fig:B}
\end{figure} 


\begin{thebibliography}{10}
\bibitem{foot:part12}
Details of the calculation presented in \Sec{sec:basic} will be reported
  separately in a broader context. For closely related calculations, see
  \Refs{my:covar, my:qdirac, ref:friedland87}.
\bibitem{foot:lz}
Of course, this is only an order-of-magnitude estimate, and $\dot{s} \lesssim
  1$ may be tolerable for specific $s$.
\end{thebibliography}

\end{document}